\RequirePackage{fix-cm}
\documentclass[twocolumn]{svjour3}          
\smartqed  
\usepackage{graphicx}
 \journalname{Journal of Geodesy}
\begin{document}

\title{Ocean calibration approach to correcting for spurious accelerations for data from the GRACE and GRACE Follow-On missions
}

\titlerunning{Ocean calibration approach for the GRACE and GRACE Follow-On missions}        

\author{Peter L. Bender        \and
        Casey R. Betts 
}


\institute{P.~L. Bender \at
              JILA, University of Colorado Boulder and NIST, UCB 440, Boulder, Colorado, USA\\
              Tel.: 303-492-6793\\
              Fax: 303-492-5235\\
              \email{pbender@jila.colorado.edu}           
           \and
           C.~R. Betts \at
            JILA, University of Colorado Boulder and NIST, UCB 440, Boulder, Colorado, USA
}

\date{Received: date / Accepted: date}

\maketitle

\begin{abstract}
The GRACE mission has been providing valuable new information on
time variations in the Earth's gravity field since 2002.  In addition, the 
GRACE Follow-On mission is scheduled to be flown soon after the end of life of 
the GRACE mission in order to minimize the loss of valuable data on the 
Earth's gravity field changes.  In view of the major benefits to hydrology 
and oceanography, as well as to other fields, it is desirable to investigate 
the fundamental limits to monitoring the time variations in the Earth's 
gravity field during GRACE-type missions.  A simplified model is presented in 
this paper for making estimates of the effect of differential spurious 
accelerations of the satellites during times when four successive revolutions 
cross the Pacific Ocean.  The analysis approach discussed is to make use of 
changes in the satellite separation observed during passages across low 
latitude regions of the Pacific and of other oceans to correct for spurious 
accelerations of the satellites.  The low latitude regions of the Pacific and 
of other oceans are the extended regions where the a priori uncertainties in 
the time variations of the geopotential heights due to mass distribution 
changes are known best.  In addition, advantage can be taken of the repeated 
crossings of the South Pole and the North Pole, since the uncertainties in 
changes in the geopotential heights at the poles during the time required for 
four orbit revolutions are likely to be small.

\end{abstract}

\section{Introduction}
\label{intro}
In 2002 the Gravity Recovery and Climate Experiment (GRACE) mission was
launched by NASA and the German space agency DLR (Tapley et al. 2004, 2013).  It is 
based on microwave measurements of changes in the roughly 200 km separation 
between two satellites on the same nearly polar orbit and flying at about 
500 km altitude.  This mission has greatly 
improved our capability for determining the Earth's gravity field and its 
changes with time. Also, in 2009 the Gravity field and steady-state Ocean
Circulation Explorer (GOCE) mission was launched by ESA (Floberghagen et al. 2011), and a large amount 
of data from it on the short wavelength structure of the Earth's geopotential 
at satellite altitudes also is now available (Gerlach and Rummel 2013). 

The first priority for measurements of time variations in the Earth's
gravity field after the GRACE mission is to fly an improved GRACE-
type mission (GRACE Follow-On) with as small a time gap as possible after the 
end of the GRACE mission (Watkins et al. 2013).  To achieve this, it wasn't 
possible for the GRACE Follow-On mission to be drag-free, although the 
mission will include laser interferometry between the two satellites in order 
to improve the accuracy for monitoring changes in the separation.  Other 
improvements in the satellite design will reduce sources of episodic spurious 
accelerations of the satellites.  However, the results over one-revolution or 
longer arcs are still likely to be limited substantially by incomplete 
corrections for noise from the on-board accelerometers that are used to 
correct for spurious non-gravitational forces on the satellites.

\begin{figure}
\includegraphics[width=0.5\textwidth]{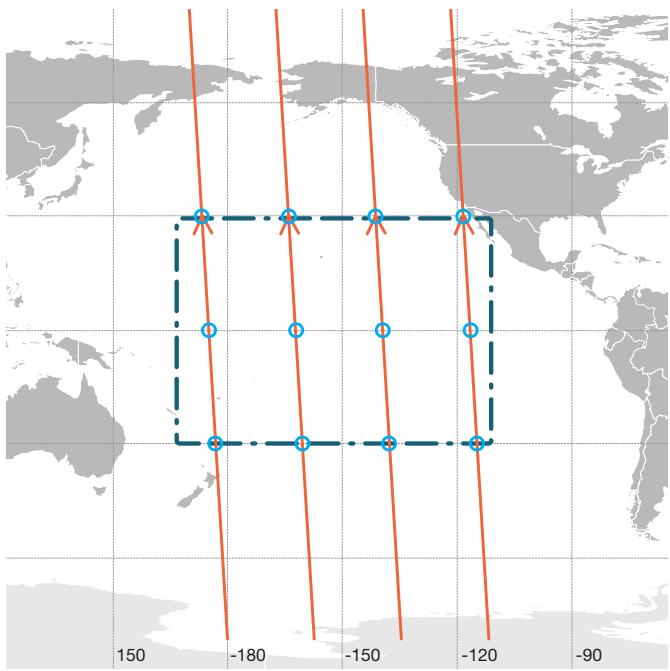}
\caption{Groundtracks for four successive upward passes of the GRACE or
GRACE Follow-On satellites across the Pacific Ocean. The circles show the locations of the assumed calibration sites in the Pacific.}
\label{fig:1}
\end{figure}

At present, for GRACE data, empirical parameter corrections are used to
reduce the effects of spurious accelerations.  A typical procedure is to
solve for once-per-revolution (once/rev) corrections to the along-track differential acceleration
from each one revolution arc of data, plus a few other parameters such as the 
mean differential acceleration.  However, a priori information on the
geopotential height variations along the orbit at satellite altitude have to
be used in solving for the parameters in the spurious acceleration
corrections.  This means that errors in the a priori geopotential height
variation models will result in errors in the final GRACE results.  In
addition, it appears possible that the empirical parameter set solved for
with present GRACE data is not sufficient to take out most of the effect of
differential spurious accelerations.

An alternate approach that has been suggested {(Bender et al. 2011, Bender and Betts 2013)} is to make use of a priori information 
on geopotential height variations only over regions where such information is 
known substantially better than at average worldwide locations.  The main 
candidate for an extended region with good a priori information is the low 
latitude region of the central Pacific Ocean. In particular, there is one 
period per day when four successive revolutions of the satellite pair will 
pass upward across the main part of the Pacific, as shown in Figure \ref{fig:1}, and 
then another period when four successive revolutions will pass downward 
across the region.  Thus the possibility of using mainly data over that 
region in correcting for spurious accelerations appears to be quite 
attractive.  However, including the data from two or more passes across low 
latitude regions of the Atlantic and Indian Oceans during the same time 
intervals would be desirable.  The sites from which the necessary data are 
obtained will be referred to as calibration sites.

To keep the situation as simple as possible, the discussion here will be 
limited to such four revolution arcs of data, which are assumed to start and 
end at the South Pole.  Use also can be made of the fact that measurements 
can be made five times at the South Pole and four times at the North Pole 
during the chosen four revolution arcs for a polar orbiting satellite pair.  
At an altitude of about 500 km, this takes only 6.3 hours, so the 
geopotential heights at satellite altitude at the poles are likely to have 
changed little during this time.  Thus multiple measurements at the poles can 
help in interpolating between low-latitude measurements over the oceans, even 
if the mean geopotential heights at the poles during the four revolutions are
not accurately known a priori.

In discussing the time variations in the geopotential $V(t)$, it is useful 
to have a measure of the changes that has the units of distance.  For this
reason, the phrase ``variations in geopotential height" will be used to 
describe those changes.  These variations $h(t)$ are defined to be
\begin{equation}\label{eq1}
	h(t) = \frac{\delta V(t)}{g},		
 \end{equation}
where $\delta V(t)$ is the difference of $V(t)$ from a model or constant value and 
$g$ is the local acceleration of gravity (Jekeli 1999).

\section{General Approach}
\label{sec:2}

A well-known problem in the analysis of GRACE data is the tendency for striping to occur in the results, if 
precautions against this are not taken. Striping can be described as a correlation of the errors along meridians, so 
that maxima and minima in the errors form ridges in the north-south direction.  It will be assumed here that a
substantial part of the source of this problem is the difficulty of knowing the systematic errors in the satellite
separation at frequencies near one cycle per revolution and at other low frequencies.  Such once/rev and low-
frequency variations can be caused by differential errors in the initial conditions for the two satellite orbits, and are
continuously modified by spurious acceleration errors.  The level to which these errors need to be known to be
useful is far better than the accuracy achievable by GPS tracking.  Thus variations at once/rev normally are at least
partially filtered out from the data.

If the mission has a particular orbit and the results are analyzed, the variations in the error at once/rev and at other
low frequencies during that period will lead to changing vertical displacements of the arcs of apparent geopotential
heights with respect to the Earth's center of mass. This will cause correlated errors along roughly the north-south
direction in the resulting gravity field solution.  Thus, it appears that providing a way to improve our knowledge of
the systematic errors in separation at once/rev and at other low frequencies, and of the spurious accelerations
leading to them, would be highly desirable. Testing the suggested procedure for accomplishing this is the
main objective of this paper.

In investigating this problem, three major approximations will be used.  One is the assumption of a very nearly spherical geopotential, with perturbations only because of short period fluctuations in the surface mass distribution.
   The second is the neglect of tidal effects.  And the third is the energy conservation approximation.
   
In the energy conservation approximation (Jekeli 1999), the separation $S$ between the satellites decreases
 whenever they fly over a region where the geopotential height near the satellite altitude is increased.  The usual
statement is that the sum of the potential energy and the kinetic energy has to stay constant, to the extent that
dissipative forces can be accounted for, so that for small changes in the potential energy, the velocity change will
be nearly proportional to the potential change.  Thus, if the potential were nearly spheroidal but with a bump in
it, each of the satellites would have about the same decrease and then increase in velocity when it crossed the bump, resulting in a
decrease in separation of the satellites.  Thus, in principle, the changes in the
geopotential height can be solved for from measurements of changes in the separation, provided that
differences in the orbit parameters and uncertainties in other factors can be solved for to the extent necessary.

From Eq. (1), the energy conservation approximation can be written as
\begin{equation}\label{eq2}
  (\bar v)\left [w_2-w_1\right ] + g*[h_2-h_1] = 0, 						
\end{equation}
where $(\bar v)$ is the mean velocity over the orbit, $w_2$ is the along-track component of the velocity for satellite 2, $g$ is the acceleration of gravity at the satellite altitude, $h_2$ is the geopotential height of satellite 2, etc.  The main amplitudes of the geophysical time variations in the geopotential height are at degree 10 or less, so to a good approximation
\begin{equation}\label{eq3}
 [h_2 - h_1] = \left [\frac{(\bar S)}{(\bar v)}\right ]*\frac{dh}{dt}                                  
 \end{equation}
where $h$ is the geopotential height at the midpoint between the two satellites.  

Inserting (3) into (2) and integrating from $t = 0$ to $t = T$ gives
\begin{equation}\label{eq4}
	(\bar v)\left [S(T)-S(0)\right ] + \left [\frac{(g*\bar S)}{(\bar v)}\right ][h(T)-h(0)] = 0.      
	\end{equation}
Eqs.(2) and (4) are clearly only approximations, and thus corrections to the 
satellite orbit parameters and offsets of the velocity vectors from the 
satellite separation vector have to be allowed for in the analysis.	
Since $g/(\bar v)^2 = 1/a$, where $a$ is the semi-major axis of the orbit, 
\begin{equation}\label{eq5}
\left [S(T) - S(0)\right ] + \left [\frac{(\bar S)}{a}\right ]\left [h(T) - h(0)\right ]=0.  
\end{equation}

If we define $\delta S$ and $\delta h$ to be the variations in $S$ and $h$ from their initial values, we have:
\begin{equation}\label{eq6}
  \delta S = -\frac{\delta h}{R},						
\end{equation}
where $R$ is given by $R = a/S_0$, $a$ is the orbit semi-major axis, 
and $R$ is about 69 for an assumed value of 
100 km for the mean separation $S_0$.  This assumed separation is about a
factor of 2 less than the nominal value for the GRACE Follow-On mission, but
the separation may be decreased if the laser interferometry operating mode
is being used.  

The geopotential height discussed throughout this paper will 
be that near the satellite altitude, for simplicity.  Thus, after results are 
obtained for a given period, they will need to be downward continued to 
obtain the results at ground level. The downward continuation can be done
by expressing the results for the geopotential at altitude in a spherical
harmonic expansion, and then reducing the amplitudes to those at the desired
reference surface.
To be specific, the polar orbit case that is used in this paper is a nearly 
circular orbit with 198 revolutions in 13 sidereal days.  The altitude is 
489 km, and there are 15.23 revolutions per sidereal day.  The offset in 
longitude between ground tracks on successive revolutions is 23.64$^{\circ}$.  
Attention will be focused on measurements of the satellite separation when 
crossing a broad region in the Pacific Ocean between 30$^{\circ}$ S and 30$^{\circ}$ 
N latitude.  The basic reason is the expected smoothness, small 
amplitude, and low relative uncertainty in the time variations in the mass 
distribution at moderate latitudes over the Pacific.  The inverted barometer 
effect would at least partially reduce the effects of uncertainties in 
surface atmospheric pressure.  And satellite altimetry results plus wind data 
derived from satellites and other sources can further reduce the uncertainty.

The orbit for GRACE does not have a fixed ground track, and neither will the orbit for
the GRACE Follow-On mission.  However, future missions after
GRACE Follow-On seem likely to have a fixed ground track, if they are drag-free.  Although such
missions probably will fly at considerably lower altitudes, it was decided to
use fixed ground track orbits in the present analysis.  For short arc
analyses, this is not expected to have any effect on the results.

The main part of this paper will be based on the expected spurious 
acceleration levels for the GRACE Follow-On mission. The main observables 
considered will be the satellite separations over 12
points in the chosen region in the Pacific, the separations at single points 
in the Indian and Atlantic Oceans, and those at the poles.  The basic set of 
measurements proposed to estimate the differential spurious accelerations 
will be described in \ref{sec:3}.  All of the latitudes referred to in the rest of this 
paper will be north latitudes, and all of the longitudes will be east 
longitudes.  

The results for the evaluation of the errors in the geopotential heights
with the ocean calibration approach will be evaluated at satellite altitude
along the orbit.  In this way, the expected accuracy of the results can be
compared with those for similar 4 revolution arcs from other approaches for
correcting for the spurious accelerations of the satellites, such as various
empirical parameter correction approaches, without the serious complication
of the limitation from temporal aliasing.  The approximation of only using
calibration measurements at 12 sites over the Pacific is not expected to
limit the accuracy substantially, but keeps the calculations considerably
simpler.

In Sections \ref{sec:4}, \ref{sec:5}, and \ref{sec:6} the limitations on the results due to two main 
causes will be discussed.  One limitation results from having to interpolate 
over periods of up to 47 minutes between the times at which measurements are 
made at the chosen calibration sites.  The other is due to uncertainties in 
the a priori information on geopotential height variations with time at the 
calibration sites.

In Section \ref{sec:4}, a specific choice of the interpolation procedure will be
discussed.  It was chosen because it does a good job of reducing the residual 
effects of the spurious acceleration noise.  It is based on a specific model 
of the spurious acceleration noise as a function of frequency that is 
intended to be close to what will be achieved during the GRACE Follow-On 
mission.  However, the nominal level of the spurious accelerations for the
GRACE mission is only a factor of three higher, so the main results are also expected
to be applicable to this case. 

In Section \ref{sec:5}, a specific, but ad hoc, model for the magnitudes of errors in 
the geopotential heights at the calibration sites will be described.  It is 
based mainly on a particular model for time variations in the ocean bottom 
pressure in the Pacific.  Then, in Section 6, the effects of the 
uncertainties from this model on the spurious acceleration correction 
proceedure will be given. 

The results for this case where the expected spurious acceleration levels 
for the GRACE Follow-On mission are assumed appear to be quite encouraging.  
Thus the possibility that a similar approach could be used for some of the 
data from the GRACE mission will be discussed in Section 7.  And finally, the
general conclusions from this study of the ocean calibration approach will be
reviewed in Section 8. 

\section{Basic Analysis Approach for Estimating Systematic Errors in the  Satellite Separation}
\label{sec:3}
	
To keep the necessary calculations as simple as possible, the conceptual 
approach discussed here is based mainly on making use of measurements of the 
satellite separations when the satellites are at $-30^{\circ}$, 0$^{\circ}$, and $+30^{\circ}$ 
latitude over the Pacific and when they are at the South Pole or the North
Pole during the same four-revolution arc.  With the assumed 13 sidereal day 
repeat satellite orbit, there will be four successive upward passes of the 
satellites from $-30^{\circ}$ to $+30^{\circ}$ latitude each day that will cross the equator 
in the Pacific Ocean between the longitudes of about 150$^{\circ}$ and 245$^{\circ}$.  These 
passes will be followed about six hours later by four downward passes from 
$+30^{\circ}$ lat to $-30^{\circ}$ lat, where the equator crossings are in the same range of 
longitudes.  Thus the major portions of these passes will be fairly well away 
from the regions of strong western or eastern boundary currents.

In addition, recent studies of uncertainties in geopotential heights over 
the oceans indicate that the uncertainties over the equatorial region of the
Indian Ocean and a substantial range of latitudes in the Atlantic are similar
to those in the equatorial Pacific (see Figure 1 in Quinn and Ponte 2011).  From the geometry, it is possible to
include two additional measurements near the equator over these oceans on
most data arcs, one in the Indian Ocean on the first of the four revolutions 
and a second measurement during the last revolution over the Atlantic.  Such
additional data will be represented here by assuming additional calibration 
points when crossing the equator during the first revolution in the Indian 
Ocean and during the fourth revolution when crossing the Atlantic.

Finally, as discussed earlier, measurements at the South Pole and at the 
North Pole during the same 4-revolution arc will be included.  This gives a 
total of 23 calibration points during each four-revolution data arc, 
including 12 in the Pacific, 1 each in the Indian and Atlantic Oceans, 5 at the
South Pole, and 4 at the North Pole.

If there were no uncertainty in the geopotential height variations at the
calibration points in the oceans or at the South or North Pole during the four 
revolution period, the variations in the satellite separation $S$ due to the 
spurious accelerations could be interpolated from the measured separations at 
the reference locations.  However, the choice of what type of interpolation
function to use requires some care because of the along-track spatial gaps of 
up to 180$^{\circ}$between some of the measurements.  Because 23 measurements 
at the calibration points are included during the four revolutions, up to 23 
basis functions could be used to fit the reference point data.  However, it 
has been found in this study that least squares fitting on a substantially lower 
number of basis functions works better.  

The reason for this result is that there actually will be differential
noise in the knowledge of geopotential height variations at calibration 
points in the Pacific that are separated by fairly short distances, down to 
below 21$^{\circ}$.  To fit such variations, basis functions with quite short 
wavelengths are needed, and they can amplify the effects of the geopotential 
height uncertainties when applied to the regions where substantial 
interpolation is needed.  Thus the number of basis functions used generally 
has been limited to about two-thirds of the number of calibration points.  

For convenience, the locations of the reference points are given in terms 
of the angular motion along track of the satellites with respect to the South
Pole crossing at the center of the four revolution arc of data.  Thus they
range from $-720^{\circ}$ to $+720^{\circ}$.  The crossings of $-30^{\circ}$, 0$^{\circ}$, and $+30^{\circ}$
latitudes in the Pacific for data sets with upward passes there will occur 
at $-660^{\circ}$, $-630^{\circ}$, $-600^{\circ}$ during the first revolution, etc.  

The first error source to be considered is the result of incomplete
correction for the differential spurious acceleration noise between the two
satellites.  It is assumed that equal weight will be given to the difference 
between the observed value and the value of the geopotential height at satellite altitude that is used at each
of the calibration sites.  Then, if the a priori value at the $i$-th calibration site is $Y_i$, an approximation function for
the $Y_i$ can be defined in terms of M basis functions $X_j(\theta_i)$ by
\begin{equation}\label{eq7}
	\vec Y(\theta) = \Sigma_j {(a_j)[X_j(\theta)]}.			
 \end{equation}
If a trial set of M basis functions is chosen, plus a set of $N$ calibration 
points, the best fitting set of coefficients $a_j$ can be chosen by the least
squares approach.  The function $\vec Y(\theta)$ that results from least squares
fitting the coefficients $a_j$ will be called the correction function, or the
interpolation function.

The starting point for the analysis is a $N\times M$ matrix {\bf A}, called the \lq\lq Ädesign 
matrix".  Here A$(i,j)$ is the value of the $j$-th basis function at the $i$-th 
calibration point.  If a vector $\vec {Y}$ of the apparent geopotential height errors $Y_i$ at the 
$i$-th measurement site is assumed, and ${\bf\rm H} = [{\bf\rm A}^T]*A$, then the vector 
$\vec {Z}$ of least 
squares fitted coefficients of the basis functions is given by
\begin{equation}\label{eq8}
	{\vec  Z}=[{\bf H}^{-1}]*[{\bf A}^T]*\vec {Y} ={\bf  K}*\vec {Y},                               
\end{equation}
where this equation defines the matrix {\bf K}.  The asterisk represents vector or matrix
multiplication.

In order to compare different 
possible choices of the basis functions, a specific criterion is needed.  To 
keep the calculations fairly simple, the criterion used here is based on just 
considering the residual noise after correction at the times during the four 
revolutions when the satellites are on the opposite side of the orbit from 
the Pacific, at latitudes of 60$^{\circ}$, 30$^{\circ}$, $-30^{\circ}$, and $-60^{\circ}$.  This set of points 
will be called the evaluation points, and the results for the specific choice 
of the basis functions that has been used will be discussed in Section 4.
     
\section{Imperfect Interpolation of the Satellite Separation Noise}
\label{sec:4}

\subsection{Basic Approach}
\label{sec:4.1}
The results of the choice of basis functions clearly depend on the spectral 
amplitude of the noise in the satellite separation due to the spurious 
accelerations.  It will be assumed here that the rms acceleration noise for 
each satellite in the Grace Follow-On mission is $3.3\times10^{-11}$ m/s$^2 /\sqrt{{\rm Hz}}$ (Foulon et al. 2013) in both the along-track and radial directions from 0.005 Hz to 0.1 Hz, and that it increases as $[0.005/f]^{0.5}$ at lower frequencies.  With these assumptions, 
the resulting noise in the along-track separation of the satellites can be 
calculated by integrating twice with respect to time, and then correcting for
the resonance with the orbital frequency at 1 cycle per revolution (cy/rev).

 \begin{table*}
 \centering
\caption{Spurious acceleration noise estimates for the GRACE Follow-On mission}
\label{tab:1}       
  
\begin{tabular}{ccrrr}
\hline\noalign{\smallskip}
Nominal & Sep. Noise & Resonance & Total & Equivalent\\
Frequency & Amp.  Without & Factor & Sep. Noise & Geopotential\\
(cy/rev) & Res. Factor & & Amplitude & Height Error\\
 & ($1\times 10^{-6}$ m)  &  & ($1\times10^{-6}$ m) & (mm)\\
 \noalign{\smallskip}\hline\noalign{\smallskip}
3/64         &2810           &3.0          &8430            &582\\
1/8           &723            &3.1          &2240             &156\\
1/4           &75.3          &3.3            &249              &17.1\\
29/64       &27.9         &4.6            &129              &9.00\\
5/8           &4.62         &5.9             &27.3            &1.89\\
11/16        &3.63        & 7.9             &28.8            &1.98\\
3/4            &2.92        &9.9            &29.1            &2.01\\
13/16        &2.39       &11.0             &26.4            &1.83\\
7/8            &1.99       &12.1             &24.0            &1.65\\
15/16         &1.67       &12.7             &21.3            &1.47\\
1               &1.42       &12.9             &18.3            &1.26\\
17/16         &1.22       &12.7            &15.6            &1.08\\
9/8             &1.06       &12.1             &12.9           &0.90\\
19/16          &0.924       &11.0            &10.2            &0.69\\
5/4             &0.813        &9.9             &8.10            &0.558\\
21/16          &0.720        &8.1             &5.82           &0.402\\
11/8           &0.639        &6.3             &4.02           &0.276\\
23/16          &0.573        &5.6            &3.21           &0.222\\
3/2             &0.516        &4.8             &2.46           &0.171\\
25/16          &0.465        &4.4             &2.04           &0.141\\
13/8           &0.420        &4.0             &1.68           &0.117\\
27/16          &0.384        &3.7             &1.41           &0.096\\
7/4             &0.351        &3.4             &1.20           &0.084\\
29/16          &0.321       &3.2             &1.02          &0.069\\
15/8            &0.294       &3.0             &0.87           &0.060\\
31/16           &0.27      &2.8             &0.75           &0.051\\
2               &0.252        &2.6             &0.66           &0.045\\
\noalign{\smallskip}\hline
\end{tabular}
\end{table*}

With these assumptions, the resulting noise in the along-track separation
of the satellites without including the effect of the orbital resonance is 
given in the second column of Table 1. These values are for the root mean square noise amplitudes for 27 frequency 
intervals centered on the nominal frequencies given in column 1.  They range 
from 3/64 to 2 cy/rev.  The bandwidths are 1/32, 1/8, 1/8, and 
9/32 cy/rev for the lowest four frequencies, and are 1/16 cy/rev for the rest.  
The sum of the frequency intervals chosen covers the range from 1/32 to 65/32 
cy/rev uniformly.  
   
Rough estimates of the effect of the orbital resonance on the satellite 
separation were obtained as follows.  The Hill equations 
(Kaplan 1976) were used to calculate the effect of a given low level of 
along-track perturbing force on one of the two GRACE satellites on the range 
between the satellites during four revolutions.  This was done for different 
perturbing frequencies, and the ratio of the rms range change to that for 
frequencies way above the orbital frequency was taken as the resonance factor 
for the along-track perturbations.  Then the same was done for radial 
perturbing forces.  The rss (root sum square) combination of these factors is listed as the 
resonance factor in column 3 of Table 1.

Finally, the error in the geopotential 
height at satellite altitude that would result from this separation noise 
level is obtained by multiplying by a factor 69, as discussed in Section \ref{sec:2}, and 
is given in the fifth column.  These rms amplitudes are called $q(k)$, where $k$ 
is the index number for the $k$-th noise frequency.  The amplitudes are large 
at frequencies below 5/8 cy/rev, but with suitable choices of basis
functions, it was found that the contributions to the interpolation errors 
for the different frequency intervals could be made to decrease at the lowest 
frequencies.

The corresponding sine and cosine functions used to represent the range noise at specific frequencies will be called $ws_k(t)$
and $wc_k(t)$, with $ws_1(t)$ and $wc_1(t)$ being the sine and cosine functions for 
the first noise frequency, $ws_2(t)$ and $wc_2(t)$ those for the second noise 
frequency, etc.  The maximum amplitudes of the noise functions are 1, except 
for the sines of any frequency less than 1/8 cy/rev.  The values of the noise 
functions at the calibration points are given by matrices {\bf BS} and {\bf BC}, where 
${\rm BS}(i,k)$ is the value of the $k$-th sine noise function at the $i$-th calibration 
point and ${\rm BC}(i,k)$ is the value of the $k$-th cosine noise function at the $i$-th 
calibration point.

Let $\bf{CS} = {\bf K}*{\bf BS}$ and ${\bf CC} = {\bf K}*{\bf BC}$.  Then CS$(i,k)$ and CC$(i,k)$ are the amplitudes of the $i$-th basis function coefficients due to unit amplitude for the $k$-th
sine and cosine noise functions.  
Also, let J$(m,i)$ be the amplitude of the correction function at the $m$-th evaluation 
site due to a unit coefficient for the $i$-th basis function.  
And, let the matrices ${\bf DS}(m,k)$ and ${\bf DC}(m,k)$ be the
values at the $m$-th evaluation site due to unit amplitude for the $k$-th sine
and cosine noise functions.  Then, if the matrices ${\bf ES} = {\bf J}*{\bf CS} - {\bf DS}$ and 
$\bf{EC} = {\bf J}*{\bf CC} - {\bf DC}$, then ${\bf ES}(m,k)$ is the error in the correction function at the $m$-th evaluation site due to a unit error in the $k$-th sine noise function, etc.

Let $k = 1$ to 4 correspond to the evaluation points at 
along-track angular locations of $-510^{\circ}$, $-480^{\circ}$, $-420^{\circ}$, and $-390^{\circ}$ with respect 
to the mid-point of the four-revolution arc, $k = 5$ to 8 correspond to $-150^{\circ}$,
$-120^{\circ}$, $-60^{\circ}$, and $-30^{\circ}$, etc. The different realizations of the noise function for different sets of four
revolutions on different days can be evaluated by randomizing the signs of
the amplitude coefficients $q(k)$ of the noise functions.  If the values of the $q(k)$ 
are those given in Table 1, then the square of the error in the correction function at the $m^{\rm th}$ evaluation point due to the rms error $q(k)$ in both the sine and cosine noise functions for the 
$k^{\rm th}$ noise frequency is given by
 
\begin{equation}\label{eq9}
	{\rm E}(m,k) = [q(k)*{\rm ES}(m,k)]^2 + [q(k)*{\rm EC}(m,k)]^2.             
	\end{equation}
	
For a given set of basis functions,
E$(m,k)$ contains all of the information needed to evaluate how well the
resulting least squares fit correction function will do in minimizing the
mean square errors at the 16 chosen evaluation sites in the hemisphere
opposite to the Pacific.  

\subsection{Results}
\label{sec:4.2}
\begin{table*}
\centering
\caption{Mean square interpolation error at the 16 evaluation sites}
\label{tab:2}       
\begin{tabular}{l*{16}{c}}
\hline\noalign{\smallskip}
    Eval. site \#   &1     &2    &3    &4   &5      &6     &7      &8 &9    &10    &11    &12    &13     &14   &15 &16\\
    Error (mm$^2$) &0.549  &0.234 &0.261 &0.738 &2.05 &2.91 &2.41&1.27 &0.855&1.85&2.50  &1.66&0.963  &0.459  &0.810 &1.63\\ 
   \noalign{\smallskip}\hline
\end{tabular}
\end{table*}

The investigation of different choices of basis functions for the
interpolation process consisted of calculating the matrix E$(m,k)$, and looking
at the amplitudes in its different columns.  From these amplitudes, the
contributions to the geopotential height uncertainty due to the interpolation 
process can be seen for each frequency interval separately.  Thus changes in 
the choice of basis functions can be considered, based on what frequency 
regions appear to need better coverage.

With the estimates of amplitude coefficients $q(k)$ given in the fifth column
of Table 1, several different combinations of up to 16 basis functions were 
chosen to try to fit the noise at the calibration points in such a way that 
the residual noise at distant locations would be small.  As discussed
earlier, the criterion used was to minimize the residual noise at the times 
during the four revolutions when the satellites were at the 16 chosen 
evaluation points.  With this criterion, a reasonable choice of M basis 
functions was found to be as follows:  $X_1=1$ and $X_2=\theta/4 \pi$, where 
$\theta$ is the angular position with respect to the center 
of the four-revolution arc; $X_3$ to $X_8$ are the sines of 5/4, 1, 3/4, 7/16, 7/32, 
and 13/128 cy/rev; $X_9$ to $X_{14}$ are the cosines of these frequencies; $X_{15}$ is a function that is 1 at the times of crossing the South Pole and is 0 at other
times; and $X_{16}$ is similarly defined for the 
North Pole.

For this set of basis functions, the matrix E$(m,k)$ was calculated.  From
it, the values of F$(m)$ were obtained, where F$(m)$ is the sum of E$(m,k)$ over $k$.
The values of F$(m)$ are the mean square interpolation errors at the individual
evaluation sites that would be obtained if there were no errors in the
knowledge of the geopotential heights at the calibration sites.  They are 
given in Table 2. 

The average of F$(m)$ over the 16 evaluation sites is $<{\rm F} (m)> = 1.32$ mm$^2$.  
Thus the expected root mean square error in the correction function at the 
evaluation sites due to the interpolation process is 1.15 mm in the 
geopotential height at satellite altitude.

As will be seen later, this measure of the rms error due to interpolation
of the correction function for the spurious acceleration noise is quite small
compared with the effect of uncertainties in the geopotential height time
variations at the calibration sites.  However, it is of some interest
to see if a simpler set of basis functions could produce nearly as small an
error due to interpolation.  As a measure of this, the calculations were
repeated for 6 different cases.  In each, one of the six basis function
frequencies was left out, so that only 14 basis functions remained.  The
results were that in each case the overall rms interpolation error increased
by a factor of 2 or more.  Thus it appeared worthwhile to keep all 16 of the
basis functions in the remaining parts of the study.

Also of interest is how the expected mean square errors over the evaluation
sites vary with the noise frequency.  To see this, the average of E$(m,k)$ over
$m$, ${\rm G}(k) = <{\rm E}(m,k)>$, is given in Table 3 for the 27 different noise 
frequencies considered.

\begin{table*}
\centering
\caption{Mean square error in geopotential height for different noise frequency bands}
\label{tab:3}       
\begin{tabular}{l*{14}{c}}
\hline\noalign{\smallskip}
 Freq. (cy/rev)      	       &3/64   &1/8   &1/4   &29/64   &5/8    &11/16  &3/4  \\  
 Error (mm$^2$)  &5e-6   &3e-6  &3e-6 &0.001   &0.001  &0.001  &4e-18   \\ \\
 Freq. (cy/rev)               &13/16 &7/8 &15/16   &1   &17/16   &9/8   &19/16   \\  
 Error (mm$^2$)    &0.001  & 0.010 &0.003   &6e-14   &0.005  &0.271  &0.015     \\ \\
 Freq. (cy/rev)        &5/4  &21/16  &11/8  &23/16 & 3/2    & 25/16  &13/8    \\
 Error (mm$^2$) &2e-19   &0.020    &0.065  &0.135 &0.171   &0.185      &0.160    \\ \\
 Freq. (cy/rev)          &27/16  &7/4 &29/16  &15/8  &31/16  &2\\
 Error (mm$^2$)&0.112   &0.078 &0.033    &0.028   &0.016   &0.011\\
\noalign{\smallskip}\hline
\end{tabular}
\end{table*}

The error contributions for 3/4, 1, and 5/4 cy/rev are less than 
10$^{-13}$ mm$^2$, since these are three of the basis function frequencies.  Also, the 
error contributions are very small for the frequency bands from 3/64 to 
1/4 cy/rev.  The main peaks are at 7/8, 9/8, and 25/16 cy/rev, and the errors 
go down smoothly for frequencies above 25/16 cy/rev.   

At an early stage in the processing of real data, the 16 parameters in 
the correction function for a particular 4 revolution arc would be
determined.  Then, this correction to the satellite separation during the arc
would be made, before the main part of the processing.  The processing could
be done either in terms of the range or the range rate, since sufficiently
precise numerical proceedures for either approach are now available (Daras et
al. 2015). 

\section{Errors in the Geopotential Heights at the Calibration Sites}
\label{sec:5}

\subsection{Model for the Errors}
\label{sec:5.1}

As emphasized by Quinn and Ponte (2011) in a recent paper:  \lq\lq Knowledge of variability in ocean bottom pressure at periods 
$<60$ days is essential for minimizing aliasing in satellite gravity 
missions."  Among the best of the currently available detailed models for 
variations in the properties of the oceans are the models produced by the 
Estimating the Circulation and Climate of the Ocean consortium (the ECCO 
models) {(Wunsch et al. 2009)}.  Results from a particular one of these models, called
dr080, are available at the ECCO-JPL website (http://ecco.jpl.nasa.gov/ external).  
Among the results are the ocean bottom pressure variations at a grid of 
points covering almost all of the oceans.  We use these results in this paper 
to estimate the variations in the geopotential at sites in the central part 
of the Pacific. This model includes assimilation of data from the oceans.

Results for ocean bottom pressure variations in the ECCO model at a number 
of sites in different oceans where there are ocean bottom pressure gauges, as 
well as comparisons of these two measures of variations, have been given in 
Quinn and Ponte (2011).  Also included are comparisons with variations from
an alternate ocean circulation model at these sites.  These results were most 
encouraging for the North Atlantic and the eastern part of the Pacific.  Our
results in a later part of this paper will be compared with the Pacific results of Quinn 
and Ponte (2011).

The model we are using for the geopotential height variations at 500 km
altitude at the calibration sites in the Pacific has four components:  
(1) random variations at each of the sites at the calibration times;  (2) a 
uniform offset at all of the sites during the roughly 6 hour calibration 
period;  (3) a north-south gradient during this period across the area covered 
by the calibration sites;  and (4) an east-west gradient across the area 
during this period.  While this model is quite crude, we believe that it will 
give a fairly good description of how the geopotential variation 
uncertainties will affect the acceleration correction approach we are 
pursuing if the amplitudes of the four terms listed above are well chosen.

As a basis for estimating the parameters to use in the geopotential 
variation uncertainty model, it would be desirable if we could start from 
knowledge of the uncertainties at different locations over some period of 
time.  However, there are limits on how well those uncertainties are known.  
One check would be to compare the results from the ECCO-JPL model with the 
geopotential variation results from present analyses of data from the GRACE 
mission.  However, since errors in the anti-aliasing geopotential models over 
both land and oceans used in analysis of the GRACE data can affect the GRACE 
results substantially, it seems quite possible that the comparison would not 
give a useful evaluation of the actual uncertainty in the geopotential 
variation  information available from sources other than the GRACE data.

In view of the above, we decided to start by looking at the full 
geopotential variations predicted by the ECCO-JPL model.  The way in which 
this was done is described in the next section.

\subsection{Geopotential Variations at Satellite Altitude from the ECCO-JPL Model}
\label{sec:5.2}

The quantity of interest in calculating variations in the geopotential at
satellite altitude are the variations in the mass per unit area as a function
of latitude, longitude, and time.  Since short wavelength variations in the
mass are attenuated rapidly with altitude, we chose to use the variations in
bottom pressure from the ECCO-JPL model only at a grid of points separated by
about 3$^{\circ}$ in both latitude and longitude.  The points included were
most of those at 3$^{\circ}$ intervals which were ocean points between $-60^{\circ}$  and 
$+60^{\circ}$in latitude and between 135$^{\circ}$ and 285$^{\circ}$  longitude, and not east of
Mexico or Central America.  The bottom pressure results at 0 and 12 hours UT 
during December 2010 were used.

To provide a check on the values at each grid point, we first subtracted 
the mean for the month and then calculated the rms variations from the mean.  
The large majority of the rms values were less than 3.0, in cm of water, and 
grid points with values greater than this have not been included in the
calculations.  In addition, 120 grid points with values between 2.0 and 3.0
were excluded, and 14 with values less than 2.0.  These were mostly at
latitudes between $-60^{\circ}$  and $-48^{\circ}$ and 
between $+36^{\circ}$  and $+45^{\circ}$ , where the
variability appeared to be much greater than at lower latitudes.  
These points were excluded because of their
variability being substantially higher than for the typical grid points
within 35 degrees of the equator, which contributed most of the variations
according to the ECCO-JPL model, and thus not being representative of how the
model would be used in practice. This left 
about 1360 grid points to be used in the calculations.

The next step was to use the differences from the mean for the month for
all the included grid points at a given latitude to calculate directly their contribution to 
the geopotential height variation at satellite altitude at each of the 12 
calibration points.  These were chosen to be at latitudes of $-30^{\circ}$ , 0$^{\circ}$, and 
$+30^{\circ}$ , and at east longitudes of 234$^{\circ}$, 211$^{\circ}$, 188$^{\circ}$, and 165$^{\circ}$.  For actual 
satellite passes northward across the Pacific, the longitudes would be about 
2 degrees higher at $-30^{\circ}$  and 2$^{\circ}$ lower at $+30^{\circ}$ N.  
However, these offsets 
would be reversed for southward passes, and these differences are not likely 
to be large enough to affect the geopotential variation model results 
appreciably.  Also, on different days, the longitudes of the passes across 
the Pacific would be shifted by up to about 12$^{\circ}$ in either direction, 
but not including these shifts is not expected to make a substantial effect 
on the results.  Including them would make the calculations considerably more
complicated.
 
\begin{figure*}
\includegraphics[width=\textwidth]{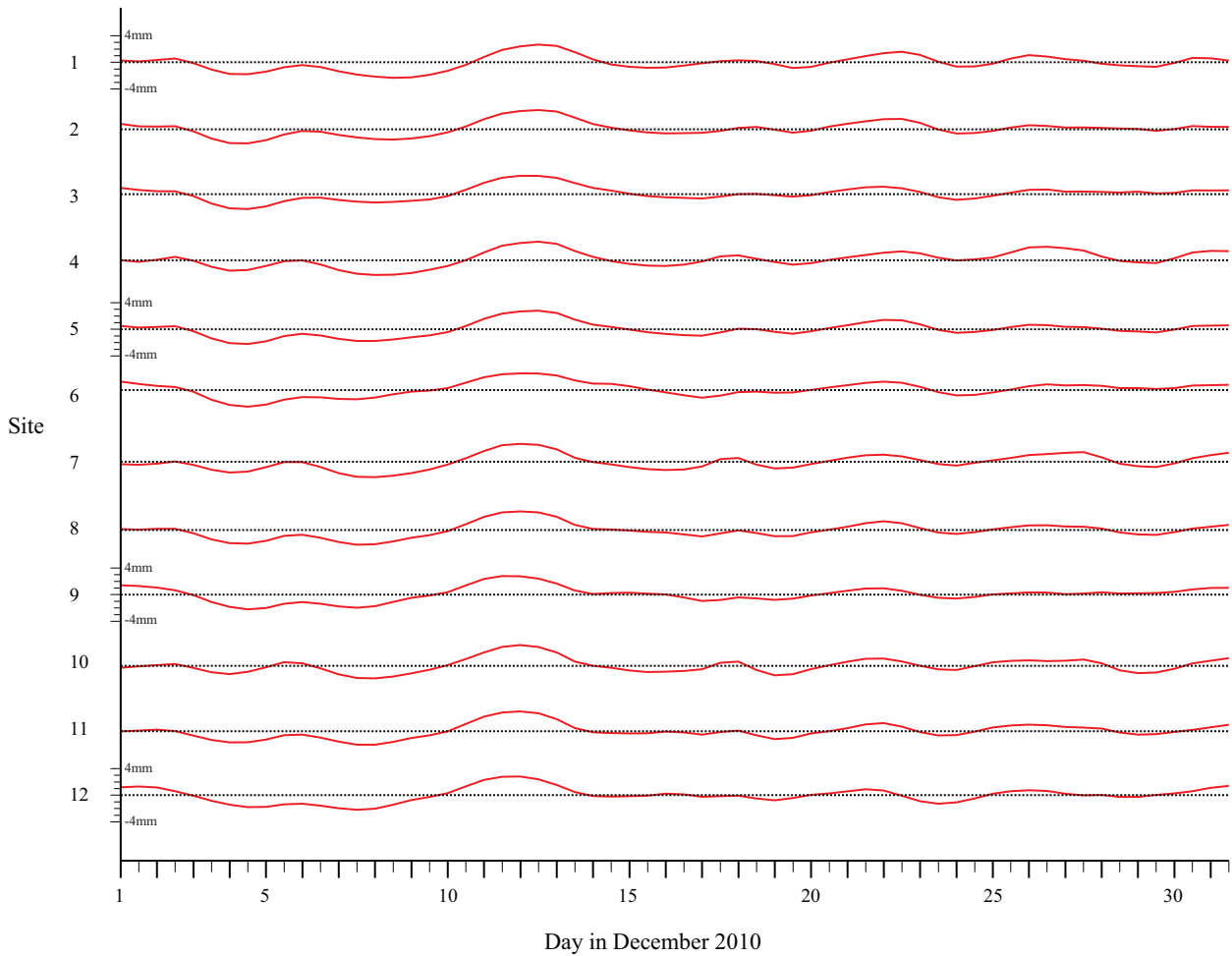}
\caption{Variations at the 12 calibration sites in the geopotential height at
satellite altitude from the ECCO-JPL ocean model.}
\label{fig:2}
\end{figure*}

 For the contributions from each latitude band of grid points and for each of the 62 times during the month, the results at the 12 calibration sites were inspected to look for outliers. 
However, the results appeared to vary quite smoothly with latitude and with time.  Thus 
the total variations from all of the latitude bands at each time were 
calculated.  This gave the values as a function of time shown in Figure \ref{fig:2}, in 
units of millimeters of geopotential height at satellite altitude.  The 
values for the times during the month and the various calibration sites range 
from $-$2.8 mm to 3.3 mm.
\subsection {Further Analysis of Geopotential Variation Results from ECCO}
\label{sec:5.3}

 In order to investigate the statistics of the variations shown in Figure \ref{fig:2}, 
we calculated the rms values over time for each calibration site.  These
values ranged from 0.99 mm to 1.28 mm.  The rms value for all of the 
calibration sites is 1.15 mm.  A comparison with the corresponding rms values 
over the whole globe for the anti-aliasing geopotential variation models used 
in analysis of the GRACE results, or the difference between two such models, 
certainly would be of interest.  It is expected that the global rms values 
would be higher, because of the larger expected uncertainties at high latitudes over the oceans and
over much of the land areas (see e.g., Thompson et al. 2004 and Gruber et al. 2011).

However, comparing with the rms values of the anti-aliasing geopotential
variation models over the rest of the globe might well not give a useful
measure of how much improvement could be obtained with the ocean calibration
approach.  The reason is that the uncertainties in the global anti-aliasing 
models are the important quantities, and these are not well known.

Other subjects of interest besides the rms values over all the calibration
sites are the variations in the average value over all 12 sites during the
month and the variations in the north-south and east-west differences across 
the Pacific calibration area across
the set of calibration sites.  The average values over all 12 sites 
range from $-$2.00 mm to 2.81 mm.  The rms of these values is 1.08 mm.  The 
variations in the average values over the 12 calibration sites are of 
considerable interest because the resulting contributions from the correlated 
variations won't be reduced in the way that the uncorrelated variations would 
be.

The variations with time in the north-south and east-west differences across the area covered by the calibration sites also can have
substantial effects.  As a measure of the north-south differences, we take the 
difference at each time between the averages of the values at the sites at 
$+30^{\circ}$ and those at $-30^{\circ}$.  These values range from $-$1.28 mm to 1.35 mm, with an rms value of 0.63 mm.  

For a rough measure of the east-west differences across the 
Pacific calibration area, we take the sum of the following: 
the average of the values at 234$^{\circ}$ minus the average of those at 165$^{\circ}$, plus
one third of the average of the values at 211$^{\circ}$ minus those at 188$^{\circ}$.  These 
values at the different times range from $-$1.09 mm to 1.50 mm.  The rms of 
these values is 0.62 mm. If the geopotential height varied linearly across the
area, this estimate would be slightly pessimistic compared with the total
geopotential height change across the $71^{\circ}$ width of the area at the
equator.

These values for the rms variations in the geopotential heights 
at the individual calibration sites, the averages over the calibration sites, 
and the differences across the calibration sites, are expected to be 
substantially larger than the errors in these quantities.  However, we have 
used them to provide an intentionally somewhat pessimistic model for the 
errors in these quantities.  If no better information on the expected model 
errors becomes available, these values can be used as the basis for a model of
the geopotential height variation uncertainties to be used in the ocean
calibration approach to correcting for spurious accelerations in future
GRACE-type missions, like the GRACE Follow-On mission, and possibly for GRACE itself.  However, before discussing this application of the results further, a partial comparison of
the ECCO model results with another source of information on ocean mass
distribution changes will be described.

\subsection{ Comparison with Ocean Bottom Pressure Gauge Results}
\label{sec:5.4}

\begin{figure*}
\includegraphics[width=\textwidth]{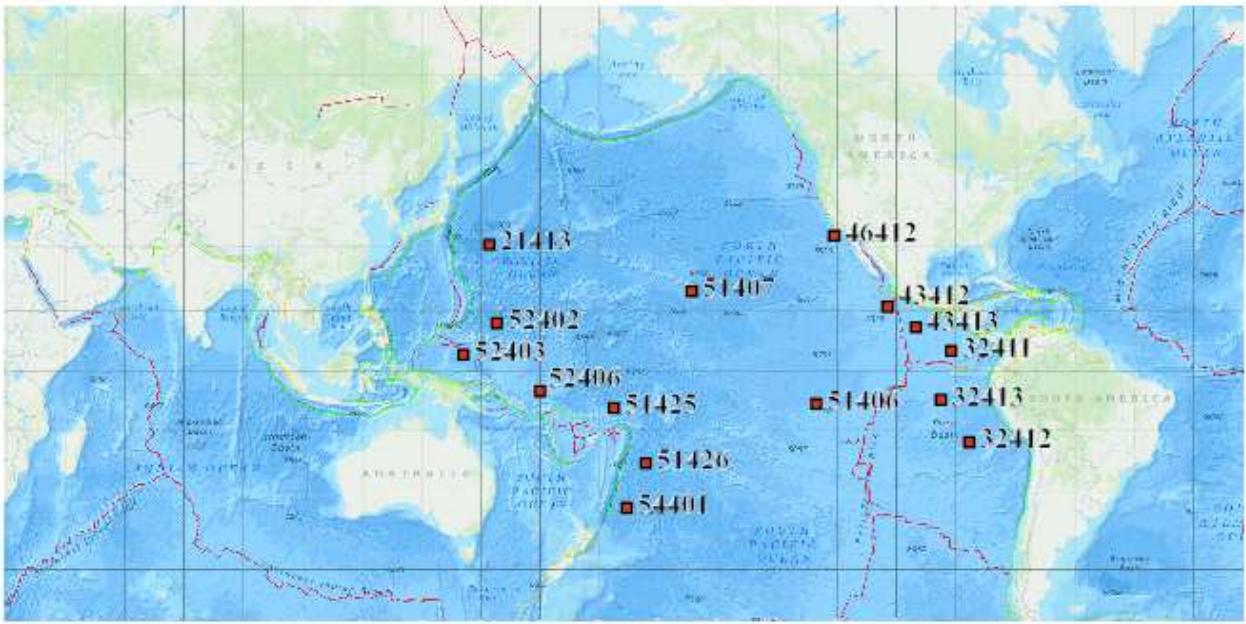}
\caption{Locations of the NOAA Deep-ocean Assessment and Reporting of Tsunamis
(DART) ocean bottom pressure gauges used in the comparison with values from
the ECCO-JPL ocean model.}
\label{fig:3}
\end{figure*}

As a partial check on the ECCO results for ocean bottom pressure 
variations, we have made a comparison with data from 15 of the ocean bottom 
pressure (BPR) gauges in the NOAA Deep-ocean Assessment and Reporting of 
Tsunamis (DART) network.  About 35 of these BPR gauges are located in the 
Pacific, with about half located fairly near the west coasts of North, 
Central, and South America plus along the Aleutian arc.  Most of the rest are 
located in the western part of the Pacific. 

Current results from the gauges are available on the website 
www.ngdc.noaa.gov/hazard/DARTData.shtml.  However, these results include the 
quite large variations due to the tides.  We fortunately were able to obtain 
results that had been corrected for the tides from the NOAA National 
Geophysical Data Center.  The data we obtained were 3 hour average values 
every 12 hours during December 2010 for 15 of the DART sites.  The 
variations in bottom pressure were then converted to variations in equivalent 
water depth, and the means for the month were subtracted.  Similar results 
were obtained for one of the ECCO grid points close to each of the DART
sites.

%
\begin{table}
\caption{Locations of the DART and ECCO sites}
\label{tab:4}
\centering
\begin{tabular}{|r|r|r||r|r|r|r|r|}
\hline
\multicolumn{3}{|c||} {DART sites}   
	&\multicolumn{4}{c|}{ECCO sites}\\[0.5ex]
\hline
\multicolumn{1}{|c|} {\raisebox{-0ex}[0pt]{Site}}   
	&\multicolumn{1}{c|}{\raisebox{-0ex}[0pt]{Lat}}    
		& \multicolumn{1}{c||}{\raisebox{-0ex}[0pt]{Long}}          
			&\multicolumn{1}{l|}{\raisebox{-0ex}[0pt]{Nor.}}   
				& \multicolumn{1}{c|}{East}  
					& \multicolumn{1}{c|}{Nor.} 
					 	& \multicolumn{1}{c|}{\raisebox{-0ex}[0pt]{East}}\\
 \multicolumn{1}{|c|} {}   
 	&\multicolumn{1}{c|}{}    
		& \multicolumn{1}{c||}{}
			&\multicolumn{1}{c|}{Lat}   
				& \multicolumn{1}{c|}{Long}& \multicolumn{1}{c|}{Lat}&	\multicolumn{1}{c|}{Long}\\
 \multicolumn{1}{|c|} {}   
 	&\multicolumn{1}{c|}{}    
		& \multicolumn{1}{c||}{}
			&\multicolumn{1}{c|}{Gr\#}   
				& \multicolumn{1}{c|}{Gr\#}&&\\[0.5ex]
\hline \hline
    21413   & 30.51   & 152.12        & 175     & 152     & 30.5     & 152.5\\
    32411    & 5.00    & 269.16         & 129     & 269     & 5.1       & 269.5\\
    32412   & -17.96  & 273.61         & 62      & 273     & -18.05   & 273.5\\
    32413   & -7.40   & 266.50          & 87      & 266     & -7.5       & 266.5\\
    43412   & 16.07   & 253.00         & 159    & 252     & 15.82     & 252.5\\
    43413   & 11.06   & 260.15        & 149    & 260     & 11.12     & 260.5\\
    46412   & 32.46   & 239.44        & 177    & 239     & 32.5      & 239.5\\
    51406   & -8.48    & 234.97        & 84     & 234     & -8.4       & 234.5\\
    51407   & 19.59   & 203.42       & 164    &  203    & 19.75   & 203.5\\
    51425   & -9.51    & 183.76         & 80     & 183     & -9.6       & 183.5\\
    51426   & -22.99  & 191.87        & 57     & 191     & -22.54   & 191.5\\
    52402   & 11.87   & 154.04        & 151    & 154    & 11.79     & 154.5\\
    52403   & 4.05     & 145.59        &126     & 145    &  4.2       & 145.5\\
    52406   & -5.29    & 165.00        & 94      & 165    & -5.4       & 165.5\\
    54401   & -33.00   & 187.02       &  47     & 187   &  -32.5    & 187.5\\ 
\hline
\end{tabular}
\end{table}

The DART sites included are shown in Figure 3.  They also are listed in 
Table 4, along with the locations of the nearby ECCO model grid sites.  The 
Dart sites ranged in latitude from $-$33.00$^{\circ}$ to $+$32.46$^{\circ}$ and in longitude 
from 145.59$^{\circ}$ to 273.61$^{\circ}$.  This range of locations was chosen to stay 
away from high latitudes and from western or eastern boundary current
regions.  DART site \#46412 is located quite close to the Southern California 
coast, but it was included because of the lack of other DART sites in the 
northeastern corner of the region of interest.

 The results of the comparisons of the DART and ECCO variations are given in
Table 5.  For each DART site, the rms values of the differences from the mean
are given for that site and for the nearby ECCO site.  In addition, the
correlation coefficient $\rho$ between the variations from the two sources is
given.
\begin{table}
\caption{Comparisons of the DART and ECCO rms water height variations}
\label{tab:5}
\centering
\begin{tabular}{|c|c|c|c|c|}
\hline
  DART   &DART $\sigma$    &ECCO $\sigma$    & Corr. \\
   Site      	       &(cm)                  & (cm)      & Coeff.\\
\hline\hline
	21413      &2.42          &2.16          &0.852 \\
	32411      &1.08          &0.75          &0.494 \\
	32412      &1.32          &0.80          &0.579 \\
	32413      &1.36          &0.62          &0.594 \\
	43412      &1.32          &0.75          &0.659 \\
	43413      &1.52          &0.62          &0.500 \\
	46412      &1.37          &0.62          &0.379 \\
	51406      &1.38          &0.74          &0.767 \\
	51407      &1.56          &0.77          &0.429 \\
	51425      &1.41          &0.57          &0.444 \\
	51426      &1.63          &0.61          &0.580 \\
	52402      &1.86          &0.86          &0.552 \\
	52403      &1.42          &0.82          &0.584 \\
	52406      &1.17          &0.62          &0.428 \\
	54401      &2.39          &1.38          &0.726 \\
\hline
\end{tabular}
\end{table}

It is interesting that the first and last of the sites listed have
substantially larger rms variations in equivalent water depth, both for the
DART and ECCO results.  The correlation coefficients at these sites have 2 of 
the 3 largest values for all the sites.  Also, 14 of the 15 sites have 
correlation coefficients above 0.40.  Thus there is a substantial
contribution to the results that is common between the two sets of
measurements. 

At essentially all of the DART sites, the rms variations in the DART
results are very roughly twice the size of the rms variations in the ECCO 
results.  This is likely to be due to a combination of the noise in the DART 
results being higher than for the ECCO model and the variations in the ocean 
bottom pressure in the ECCO model being somewhat smaller than the real 
variations.  Since the correlation coefficients between the two types of 
variation results are typically about 0.5, there is enough correlation to 
suggest that the magnitude of the actual variations is somewhere in between.  

In the paper mentioned earlier {(Quinn and Ponte 2011)}, the ECCO data were
compared with ocean bottom pressure gauge (BPR) data and with another ocean 
model at gauge locations in the Atlantic and Indian Oceans, as well as in the 
Pacific.  The other ocean model used was the Ocean Model for Circulation and 
Tides (OMCT), that has been used as a de-aliasing model for GRACE data.  They found that the variance was only about 1 cm$^2$ of equivalent 
water depth variation in the Pacific.  The correlations of BPR results with the 
ocean model results had a wide range of values, but with values of 0.3 or 
more at many sites.  The rms variations for the BPR data in the eastern 
Pacific were only 2 to 2.5 cm, which is consistent with the variations found 
in this paper.

Our results and those of Quinn and Ponte don't give direct estimates of the
uncertainty in the mass distribution variations in the ECCO model.  However,
it seems unlikely that the uncertainty is appreciably larger than the
variations in the model.  The uncertainty also could be substantially
smaller, without this being apparent from the comparisons that have been
done.  

In view of the above, we have made what we regard as somewhat conservative estimates 
of the values to use in our model for the four types of errors for the
Pacific sites included in model.  The estimated errors are as follows:  
(1) 1.5 mm for the random errors at each of the Pacific calibration sites;  
(2) 1.5 mm for the common error at the sites during the 4 revolution 
observation period;  (3) 1.0 mm for the N-S gradient;  and (4) 1.0 mm for the E-W gradient.  For (1) and (2), the value 
chosen is about 35\% higher than the rms value for the ECCO data, to allow for the
possibility of that model actually not including some of the real ocean
variability.  For (3) and (4), the values chosen are near the maximum values  
found from the ECCO data for a similar reason.

As emphasized earlier, the uncertainties in geopotential height time
variations are believed to be less over relatively low latitude regions in
the oceans than over land or at higher latitudes.  However, to evaluate the
effect of these uncertainties, some sort of model for them is needed.  What
has been done in the past is to take two different models for the variations
in the ocean mass distribution with time, and compare their results, or to
compare one model against other types of data.  The problem with comparing
different models is that they share some of their input data, and thus their
errors may be considerably correlated.  For comparisons against other types
of data, the problem is the limitation of coverage of data from other 
sources, except for comparisons of the ocean surface height variations with 
altimetry results.

As an alternate approach, it was decided to evaluate the variations in
geopotential height at satellite altitude according to one ocean model, and
then assume that the uncertainty is equal to some fixed fraction of the
variations obtained from the ocean model.  

\section{Errors due to Uncertainties in Estimation of the Geopotential Height
       Variations at the Calibration Sites}
\label{sec:6}

Although the situation would be slightly different for four revolution arcs
with downward crossings across the Pacific from those with upward crossings,
the results are expected to be nearly the same.  Thus only the upward
crossing case has been studied.  Also, the choice of measurements at only
three latitudes in the Pacific is clearly not realistic, but it is expected
that it will give a good indication of what to expect.  The errors in
geopotential height estimates are expected to be quite well correlated over
distances considerably less than 30$^{\circ}$.

Our simplified error model based on uncertainties in the geopotential 
heights at satellite altitude is summarized in Table 6.  For the sites in the 
Pacific, a correlated error of 1.5 mm is assumed for all 12 sites, plus 
random 1.5 mm errors.  In addition, an error of 1.0 mm amplitude is assumed 
for the N-S gradient between the sites at 30$^{\circ}$ N and those at 30$^{\circ}$ S, and 
a 1.0 mm error for the E-W gradient.  For the remaining sites, random errors 
of 2.5 mm are assumed for the Indian and Atlantic Ocean sites, which is 
similar to the rss of the different errors at a given Pacific Ocean site.  
And random errors of 0.5 mm for each of the four North Pole and five South Pole 
observations are included, to allow for the variations during the times 
between those observations.

A study of the variability of the geopotential height over favorable
low-latitude sites in the Atlantic and Indian Oceans based on the ECCO-JPL 
model has not been carried out.  However, from Figures 1b and 1c in Quinn and
Ponti (2011), it does not appear that the variability at such sites would be
substantially worse than at central Pacific sites.  Also, from their Figure 2b, the
difference between the OMCT ocean model and the ECCO model appears to be
fairly small for both the Atlantic and Indian Ocean sites.  Thus it seems
reasonable to adopt 2.5 mm as the estimate for the variability for the
calibration sites in these locations.

   \begin{table}
\caption{Ad hoc error model for the geopotential height variation uncertainties}
\label{tab:6}       
\begin{tabular}{llr}
\hline\noalign{\smallskip}
Pacific sites:& & \\
\ \ Uniform error at 12 sites:	 &1.5  mm\\
\ \ Random error at sites:		 &1.5  mm\\
\ \ Linear N-S variations:                &1.0 mm\\
\ \ Linear E-W variations:		   &1.0  mm\\ \\
Random errors at Indian \\
\ \ Ocean and Atlantic Ocean sites: &2.5  mm\\ \\
Random errors at North\\ 
\ \ Pole and South Pole crossings\\ 
\ \ due to time variations:              &0.5  mm\\
\noalign{\smallskip}\hline
\end{tabular}
\end{table}

To give some indication of the scale of the errors assumed in the error
model, it is useful to consider the change in geopotential height at
satellite altitude due to a Gaussian disk of water with 1 cm height at the
center and a half-mass radius of 39$^{\circ}$.  This much additional water would
increase the geopotential height by 1.0 mm, equal to two-thirds of the correlated
and the random error assumed at the Pacific sites.  However, validation of
the assumptions in the error model will require improved evaluations of the
uncertainties in present ocean circulation models and their correlations over
the distances between the chosen calibration sites, as discussed in
Section 5.1.

It should be noted that the uncertainties in the geopotential time
variations assumed here are equal to all of the time variations from the 
ECCO-JPL model.  This may be regarded as a fairly pessimistic assumption, 
since the uncertainties actually could be substantially less.  However, it is 
difficult to find evidence for the model being better than this over the time 
scales of interest.  Comparisons with ocean bottom pressure gauge results 
from the DART network, as discussed earlier, indicate that the uncertainties 
aren't likely to be substantially worse than assumed.  But the noise level in 
those data is large enough to prevent a more precise comparison from being 
made.

The final step in the evaluation of the ocean calibration approach was to
use the set of 16 basis functions discussed in Sec. \ref{sec:4.2} along with the ad hoc
geopotential height variation error model to see how large the resulting
errors in the correction function at the evaluation sites would be.  If a
particular case of the error model from Table 6 is chosen, such as the common
error of 1.5 mm at the 12 Pacific sites, this defines the vector $\vec {Y}$ of
geopotential height errors described just before eq. \ref{eq3} in Section \ref{sec:3}.  Then, 
if {\bf K} is the matrix defined in eq. \ref{eq4} and {\bf J} is the matrix defined in Section \ref{sec:4.1},
the resulting errors at the 16 evaluation sites will be given by the vector
$\vec {L}$, where 
\begin{equation}\label{eq10}
	\vec {L} = \vec{J}*\vec{K}*\vec {Y} 
\end{equation}                                            
The calculation of $\vec  {L}$ is repeated for each of the independent errors in
Table 6, and the squares of the 16 entries in each resulting vector $\vec L$ are
averaged to give the mean square error at the evaluation sites due to the
errors in the ad hoc error model.  The resulting contributions to the average 
mean square error at the evaluation sites are given in Table 7.
                                                        
    \begin{table}
\caption{Average mean square errors at the evaluation sites in mm$^2$}
\label{tab:7}       
\begin{tabular}{lr}
\hline\noalign{\smallskip}
Independent errors at the Pacific sites: &1.62\\
Common error at the Pacific sites:	&0.10\\
North-south variations:			&2.01\\
East-west variations:			&0.29\\
Indep. error at Indian site: 	         &2.74\\
Indep. error at Atlantic site:		&1.10\\
Indep. errors at Pole crossings:		&0.42	\\  
                                                       &{\underline{\hskip4em}}\\
Total (mm$^2$)                                &8.28\\
\noalign{\smallskip}\hline
\end{tabular}
\end{table}

With the contribution of 1.32 mm$^2$ from the spurious acceleration noise, this
gives a mean square error of 9.60 mm$^2$ at the evaluation sites, or an rms
error of about 3.1 mm.  If the values in the geopotential variation error 
model were half as large as assumed, the rms error at the evaluation 
sites would be reduced to about 1.8 mm. These values appear to give a reasonable
range for the estimated geopotential height uncertainties from the GRACE Follow-On 
mission, if other comparable or larger sources of low-frequency range uncertainty
do not show up in the data.

It unfortunately is true that only about half of each day would be
covered by the 4 revolution arcs that cross the central Pacific.  Thus, other
approaches to correcting for spurious accelerations or other low frequency
errors would have to be used the rest of the time.  However, these other approaches could be
compared with the results from ocean calibration during the 4 revolution arcs
that do cross, in order to possibly help in determining which of the other 
approaches to use the rest of the time.
          
\section{Possible Application to GRACE Data}
\label{sec:7}

For the GRACE mission, the acceleration noise level requirements were a 
factor of 3 less severe than for GRACE Follow-On.  Thus, the overall requirement 
on the range acceleration noise level due to the accelerometers in the two 
satellites was
\begin{equation}\label{eq11}
	{\rm PSD}^{1/2}<3\times 10^{-10}
	 \bigg [1 + \frac{0.005\ \rm Hz}{\rm f}\bigg ]^{0.5} {\rm m}/{\rm s}^2/\sqrt{{\rm Hz}}
\end{equation}     

Based on this nominal acceleration noise level for GRACE, the mean square
contribution to the geopotential height uncertainty at the evaluation sites
would be 12 mm${^2}$.  With the 8.3 mm${^2}$ contribution from the uncertainties in
the geopotential heights at the calibration sites, this gives a total mean
square uncertainty of 20.3 mm${^2}$ at the evaluation sites, or an rms
uncertainty of 4.5 mm.  This is just a factor 1.5 higher than was found
earlier for GRACE Follow-On.

In view of this result, it appears useful to consider the possibility that
the ocean calibration approach could be tested using selected subsets of the
GRACE data.  But, unfortunately, it is difficult to tell how much of a change
this approach might make, since a number of other error sources are present 
besides the nominal level of accelerometer noise.  The sources of real or 
apparent extra low-frequency range acceleration noise in the early GRACE data 
were discussed quite early by Flury et al. (2008), and in other papers that they refer
to.  One of these sources is related to the fairly frequent thruster firing 
for attitude control.  A second is short mechanical disturbances called
twangs that are triggered by temperature changes somewhere in the satellite, 
and a third is short pulses apparently due to electrical disturbances.
    
Flury et al. (2008) were able to find some periods of 70 to 300 seconds during
which none of these disturbances appeared to be present.  Based on the data
from such periods, they concluded that the different components of the
acceleration noise for the two GRACE satellites met or slightly exceeded the
requirements at frequencies of 30 to 300 mHz. Also, Figure 5 of Fromm- knecht et al. (2006) indicates only a moderate increase in the level of low-frequency range noise at frequencies down to 1 mHz. This figure was based on data free from thruster firings, but not other disturbances, in order to make use of somewhat longer data arcs.

More recently, the total low-frequency range noise based on GRACE KBRR 
range data in 2006 was considered by Ditmar et al. (2012).  The quantity 
plotted in Figure 2 in that paper is the second derivative of the KBRR range 
minus the calculated range from six-hour satellite dynamical orbits fit to the 
data.  A curve showing the expected noise due to one component of the nominal 
accelerometer noise level is included also, but without an allowance for 
resonance at 1 cycle/rev.  If a factor of about 13 is allowed for resonance, 
the actual noise near 1 cycle/rev is about 20 times higher than that due to 
the nominal total accelerometer noise.  

A substantial amount of the excess low-frequency noise may be due to
limitations in the orbit calculations and the static geopotential that were
used when the calculations were done.  However, temporal aliasing also is
believed to be a basic limitation on all analysis methods.  Since known noise
limitations prevent the ocean calibration approach from being extended beyond
favorable 4 revolution arcs, for GRACE data, other analysis methods would need to be
evaluated over similar length arcs in order to achieve a fair
comparison.  Thus a combination of the different approaches would be needed
in order to give useable monthly solutions with GRACE data. 

\section{Conclusions}
\label{sec:8}

From the results for the case of GRACE Follow-On in Section 6, the prospects
appear to be good for being able to correct for low frequency range noise to
a level of about 3 mm in the geopotential height if the uncertainties in the
geopotential heights at the calibration sites are as small as specified in
the model described in Section 5.  This model was based mainly on the
assumption that the uncertainties at ocean calibration sites would not be
larger than the rms amplitudes of the variations from the monthly means as
derived from the ECCO-JPL ocean model.  It seems plausible that the actual  
variations would be smaller than this, but there does not appear to be any
fairly firm evidence available on this question.

An equally large or larger question is the uncertainty in the geopotential
heights over the less favorable regions of the globe that would result if the
ocean calibration approach is not used.  The various sources of mass
variations have been reviewed recently by Gruber et al. (2011).  If the methods of
correcting for noise in the accelerometers that has been used for GRACE is
used for GRACE Follow-On also, models for what is known about the mass
variations from other sources have to be applied before the corrections are
made.  Thus the errors in those models will be built into the final
geopotential results.  These models are usually called anti-aliasing models,
and their accuracy unfortunately is not well known.

From Gruber et al. (2011), Figure 13, the monthly mean variations in the global
mass distribution due to hydrology are the largest ones, followed over a
substantial range of harmonic degrees by the ice signal, the atmospheric
signal, and the ocean signal.  The hydrological mass variations are quite
well known over some areas, but not over others.  And for the oceans, the
uncertainty in the mass variations at high latitudes appears to be a lot
higher than at low latitudes.  Thus it appears plausible that the
geopotential variation results obtained with the ocean calibration approach
may be more accurate than those obtained otherwise, but we don't have any
estimate of how much of an improvement might be expected.  

A number of simulations of possible future Earth gravity missions have
been carried out recently in Europe in support of the project ``Satellite
Gravimetry of the Next Generation (NGGM-D)".  The results of the project,
also referred to as the ``$e^2$.motion" project, have been published recently by
Gruber et al. (2014).  Some studies for a single pair of satellites in the
same nearly polar orbit were included, and the results for the cumulative
geoid error for this case are shown in the left part of Fig. 7-6 in the report.  

However, it is unfortunately very difficult to compare the expected results 
for these roughly 32 day simulations with what has been found for 4 
revolution arcs across the Pacific with the ocean calibration approach.  The
assumptions made are quite different, and the much longer simulations are
affected quite strongly by temporal aliasing, because of time variations in
the geoid during the time taken for the the satellite ground tracks to cover
the Earth.  On the other hand, fitting of fairly large numbers of empirical
parameters over the much longer simulation period in the $e^2$.motion studies to 
correct for various error sources may have led to the loss of some real 
geophysical variation information.  

It is hoped that matched simulations of the different approaches can be done 
soon in order to see if the ocean calibration approach really would produce 
improved results for those 4 revolution arcs of GRACE Follow-On data that 
cross the equatorial Pacific.  Also, for future missions after GRACE 
Follow-On that are drag-free and have somewhat reduced acceleration noise at 
low frequencies, it appears possible to extend the ocean calibration approach 
to 12 revolution arcs where the middle 4 revolutions don't cross the Pacific.
Such solutions could be combined to give continuous results on the variations 
in the Earth's mass distribution. 

\begin{acknowledgements}It is a pleasure to thank the many people who have contributed to discussions of the accuracy limitations for GRACE-type missions, including
particularly David Wiese, John Wahr, Steve Nerem, Oscar Colombo, Jakob Flury, Pieter Visser,
and Helen Quinn. We would also like to thank George Mungov of the NOAA National Geograpical Data  Center for providing the tidally corrected DART ocean bottom pressure results.
\end{acknowledgements}


\end{document}